\documentclass[journal]{IEEEtran}
\usepackage{lipsum}
\usepackage{multicol}
\usepackage[utf8]{inputenc}
\usepackage{fourier} 
\usepackage{array}
\usepackage{wrapfig}
\usepackage{makecell}
\usepackage{color,soul}
\usepackage{bm}
\usepackage{cite}
\usepackage[cmex10]{amsmath}
\usepackage{array}
\usepackage{url}
\usepackage{graphicx}
\usepackage{atbegshi}% http://ctan.org/pkg/atbegshi
\AtBeginDocument{\AtBeginShipoutNext{\AtBeginShipoutDiscard}}
\usepackage{caption}
\usepackage{subcaption}
\usepackage{algorithm}
\usepackage{algpseudocode}
\usepackage{epstopdf}
\usepackage{multirow}

\usepackage[font={small}]{caption}
\usepackage[justification=centering]{caption}
\usepackage{amsmath}
\makeatletter
\renewcommand{\ALG@beginalgorithmic}{\small}
\makeatother
\makeatletter 
\def\@eqnnum{{\normalsize \normalcolor (\theequation)}} 
\makeatother
\usepackage{cleveref}
\makeatother
\usepackage{graphicx}
\usepackage{xcolor}
\usepackage{scalerel}
\usepackage{color,soul}
\usepackage{color}
\usepackage{empheq}
\usepackage[]{graphicx}
\usepackage{epstopdf}
\usepackage{empheq}
\usepackage{array}
\usepackage{booktabs}
\usepackage[cmex10]{amsmath}
\usepackage{array}
\makeatletter
\newcommand*{\rom}[1]{\expandafter\@slowromancap\romannumeral #1@}
\makeatother

\begin{document}
% paper title
% can use linebreaks \\ within to get better formatting as desired
\title{Multi-Terminal DC Fault Identification for MMC-HVDC Systems based on Modal Analysis - A Localized Protection Scheme}
% author names and IEEE memberships
% note positions of commas and nonbreaking spaces ( ~ ) LaTeX will not break
% a structure at a ~ so this keeps an author's name from being broken across
% two lines.
% use \thanks{} to gain access to the first footnote area
% a separate \thanks must be used for each paragraph as LaTeX2e's \thanks
% was not built to handle multiple paragraphs
\author{Vaibhav~Nougain,~Sukumar~Mishra, \textit{Senior Member, IEEE},
George S. Misyris, \textit{Student Member, IEEE}
and~Spyros Chatzivasileiadis, \textit{Senior Member, IEEE}}

\thanks{Vaibhav Nougain and Sukumar Mishra are with the Department
of Electrical Engineering, Indian Institute of Technology, Delhi, New Delhi, 110016 India. e-mail: (nougainvaibhav@gmail.com, sukumariitdelhi@gmail.com).}
\thanks{George S. Misyris and S. Chatzivasileiadis are with the Center for
Electric Power and Energy (CEE), Technical University of Denmark (DTU),
Kgs. Lyngby, Denmark. e-mail: (gmisy@elektro.dtu.dk, spchatz@elektro.dtu.dk).}
\thanks{George S. Misyris and S. Chatzivasileiadis acknowledge the support of Innovation Fund Denmark through the project "multiDC", Grant No. 6154-00020B.}
\maketitle

\begin{abstract}
%\boldmath
We propose a localized protection scheme based on modal analysis in multi-terminal modular multilevel converter (MMC) based high voltage DC (HVDC) systems. The paper addresses the issues of localized protection scheme based DC fault identification, such as differentiating between external and internal faults, classification of type of fault contingency i.e., pole to pole (\textit{PTP}) or pole to ground (\textit{PTG}) for high impedance faults (HIFs) in the system. The scheme works on equivalent network of multi-terminal MMC-HVDC system for a DC fault, using phase-modal transformation to analyse line-mode and zero-mode voltage across the current limiting reactor (CLR) for different possible contingencies in the presence of fault resistance. The protection scheme is validated to be reliable for HIFs and in the presence of White Gaussian Noise (WGN) in measurement. The scheme operation is validated to be intact for varying fault location, fault resistances and system transients.
\end{abstract}
% IEEEtran.cls defaults to using nonbold math in the Abstract.
% This preserves the distinction between vectors and scalars. However,
% if the journal you are submitting to favors bold math in the abstract,
% then you can use LaTeX's standard command \boldmath at the very start
% of the abstract to achieve this. Many IEEE journals frown on math
% in the abstract anyway.

% Note that keywords are not normally used for peerreview papers.

% For peer review papers, you can put extra information on the cover
% page as needed:
% \ifCLASSOPTIONpeerreview
% \begin{center} \bfseries EDICS Category: 3-BBND \end{center}
% \fi
%
% For peerreview papers, this IEEEtran command inserts a page break and
% creates the second title. It will be ignored for other modes.
\IEEEpeerreviewmaketitle

\section{Introduction}
% The very first letter is a 2 line initial drop letter followed
% by the rest of the first word in caps.
% 
% form to use if the first word consists of a single letter:
% \IEEEPARstart{A}{demo} file is ....
% 
% form to use if you need the single drop letter followed by
% normal text (unknown if ever used by IEEE):
% \IEEEPARstart{A}{}demo file is ....
% 
% Some journals put the first two words in caps:
% \IEEEPARstart{T}{his demo} file is ....
% 
% Here we have the typical use of a "T" for an initial drop letter
% and "HIS" in caps to complete the first word.
\IEEEPARstart {T}{HE} idea of better scalability and very low harmonics has made modular multilevel converter (MMC) technology effectively applicable for HVDC systems with the objective of transmitting large offshore wind energy over long distances [1]-[3]. Multi-terminal MMC-HVDC configuration ensuring system reliability with continuous operation of power transfer even under DC faults has resulted in practical offshore wind projects like the four-terminal ±500 kV/3000 MW HVDC project in Zhangbei, China [3]. However, in case of a multi-terminal configuration, since there are numerous MMC terminals finding different paths (with different impedances) to contribute to the fault current, the fault current is relatively higher compared to the conventional point to point HVDC system [4], with only two terminal contribution. In order to mitigate the damage to the power electronic switches of MMC converter due to the fault current, direct current circuit breakers (DCCB) are used to isolate the faulty part of the network in time ensuring continuous power transfer in the multi-terminal MMC-HVDC system [5]. Therefore, a rapid and reliable fault identification scheme for DC faults is an indispensable requirement for such configurations.
\newline
Considering the fault identification schemes for multi-terminal MMC-HVDC system in the literature, travelling wave (TW) based localized methods are effective and hence, popular. The authors in [6] employ rate of change of local current, synonym of ROCOC. The authors in [7] take the ROCOV of the fault limiting reactor to identify the fault upon inception. However, high impedance faults (HIFs) and the presence of white gaussian noise (WGN) jeopardize the reliability of the schemes in [6]-[7]. The rate of change of ROCOV of the current limiting reactor (CLR) has been taken as the decisive parameter by authors in [8] working on the problem of maloperation due to WGN. However, the decision on minimum time window required for noise immunisation is a tricky affair for complex systems with multiple converters [9]. The authors in [10] take the transient energy ratio of DC Filter Link as the decisive parameter for fault identification whereas the authors in [11] employ the CLR Power. The authors in [12] propose a scheme integrating rate of transient voltages and communication based backup protection. The schemes in [10]-[12] work effectively for HIFs. However, the performance of such schemes is sensitive to fault type and fault resistance, also the impact of noise [14] has also not been discussed in [10]-[12]. In addition, as pointed in [14], accurately distinguishing between external and internal faults is a problem unresolved with schemes proposed in [6]-[12].
\newline
Working on the aspect of selectivity for external and internal faults, the authors in [13] use frequency domain based ratio of transient voltages to distinguish between external and internal faults in the system. However, the scheme has weakness to WGN [14]. The authors in [14] employ phase-modal transformation to propose an elaborate analysis working on effect of WGN and distinguishing between external and internal faults. The scheme proposed in [14] gives a good theoretical foundation to classify the faults based on line-mode and zero-mode voltage across the current limiting reactor. However, the scheme does not consider fault resistance, \textit{R$_{f}$} in the analysis anywhere, realising the equivalent networks for bolted faults only. Also, the line resistance and the arm resistance are neglected in the analysis. In order to distinguish between an external forward fault and an internal fault, the authors in [14] consider the magnitude of line-mode voltage where external forward fault has lesser line-mode voltage magnitude than an internal fault. The analysis holds true for a bolted fault. However, HIFs may have severe effects on the selectivity of the scheme while distinguishing between an external forward and an internal fault. 
\newline
The scope of this paper is to offer an imperative improvement to the fault identification scheme proposed in literature in terms of selectivity while distinguishing between external and internal faults, including the effect of fault resistance, line resistance and arm resistance in the analysis. The proposed work use phase-modal transformation to analyse the equivalent network classifying different faults based on the line-mode and zero-mode voltage. The proposed scheme is reliable for HIFs and in the presence of WGN in measurement. The scheme is also selective for external faults.
\newline
The rest of the paper is organized as follows. Section \rom{2} gives the test system configuration elaborating the MMC equivalent model and the overhead line (OHL) equivalent model. Section \rom{3} gives the fault analysis for different fault contingency in the system based on line-mode and zero-mode voltage. Section \rom{4} presents the proposed fault identification algorithm. Section \rom{5} gives the validation of the scheme, presenting the results for the protection algorithm. Finally, section \rom{6} concludes the paper.
\newline
\begin{figure}[!thb] 

		\centering
	\includegraphics[width=\columnwidth,  height= 4 cm]{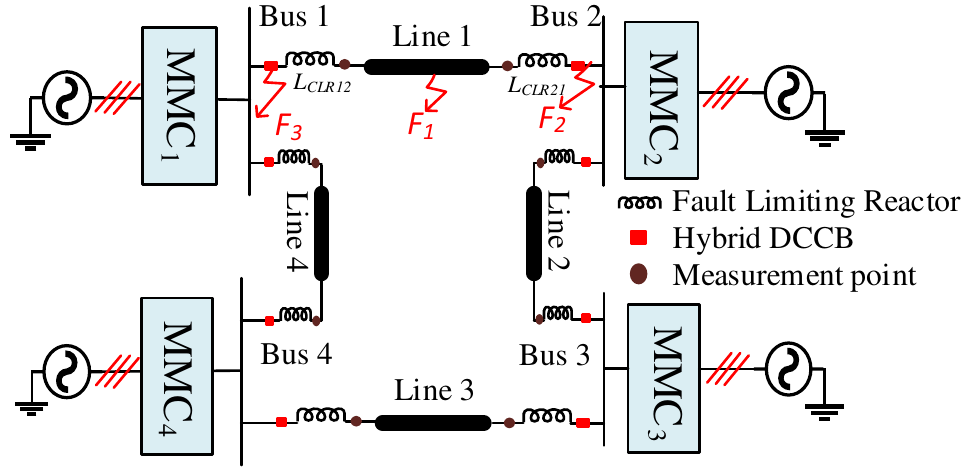}
	\caption{Four-terminal MMC-HVDC test system}
	\label{fig:sys_model}
\end{figure}
\section{Test System Configuration}
A four-terminal half-bridge MMC-HVDC configuration is taken into consideration as the test system [17] to validate the proposed fault identification scheme [refer to Fig. 1]. Current limiting reactors (\textit{L$_{CLR}$}) are employed to mitigate the rate of rise of current in case of a fault contingency in the system [18]. Fault \textit{F$_{1}$} shows a fault in line 1 at a specific location whereas fault \textit{F$_{2}$} shows a forward external fault and \textit{F$_{3}$} shows a backward external fault in the system. A symmetrical monopole configuration with frequency-dependent transmission model (FDTL) for overhead lines (OHL) is considered for the system [14].   
\subsection{Half-bridge MMC simplified equivalent model}
The half-bridge MMC-HVDC configuration is represented with its simplified equivalent model [19] as shown in Fig. 2. The parameters in the simplified model are defined as, \textit{R$_{MMC}$=2/3R}, \textit{L$_{MMC}$=2/3L}, \textit{C$_{MMC}$=6C/N}. Here \textit{R} is the arm resistance, \textit{L} is the arm inductance, \textit{C} is the sub-module (SM) capacitance and \textit{N} gives the number of SMs per arm.
\begin{figure}[!thb] 

		\centering
	\includegraphics[width=\columnwidth,  height= 2.8 cm]{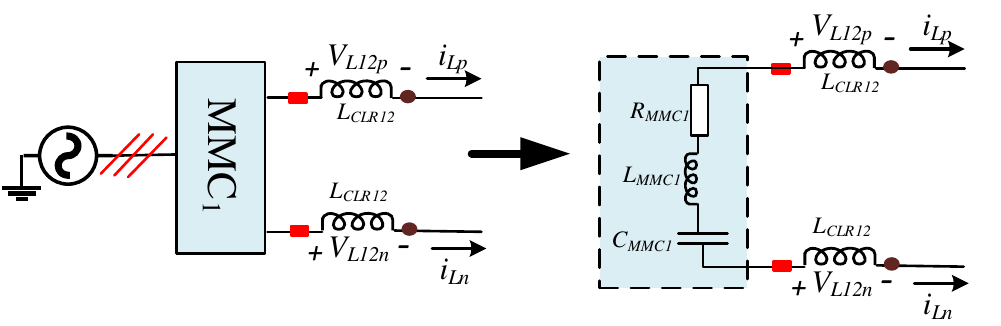}
	\caption{Simplified equivalent model of half-bridge MMC system}
	\label{fig:sys_model}
\end{figure}
\subsection{OHL equivalent model}
The grounding capacitance for OHL is within 10$^{-2}$ $\mu$F/km [20] while the equivalent DC capacitor of MMC is around 10$^{3}$-10$^{4}$ $\mu$F [21]. Hence, the fault contribution from the grounding capacitance of OHL can be ignored [22]. Therefore, OHL transmission is simplified to R-L equivalent model as shown in Fig. 3 where \textit{L$_{mn}$}, \textit{R$_{mn}$} and \textit{M$_{mn}$} are self-inductance, resistance and mutual inductance of OHL \textit{mn}. 
\begin{figure}[!thb] 
\centering
	\includegraphics[width=\columnwidth,  height= 7 cm]{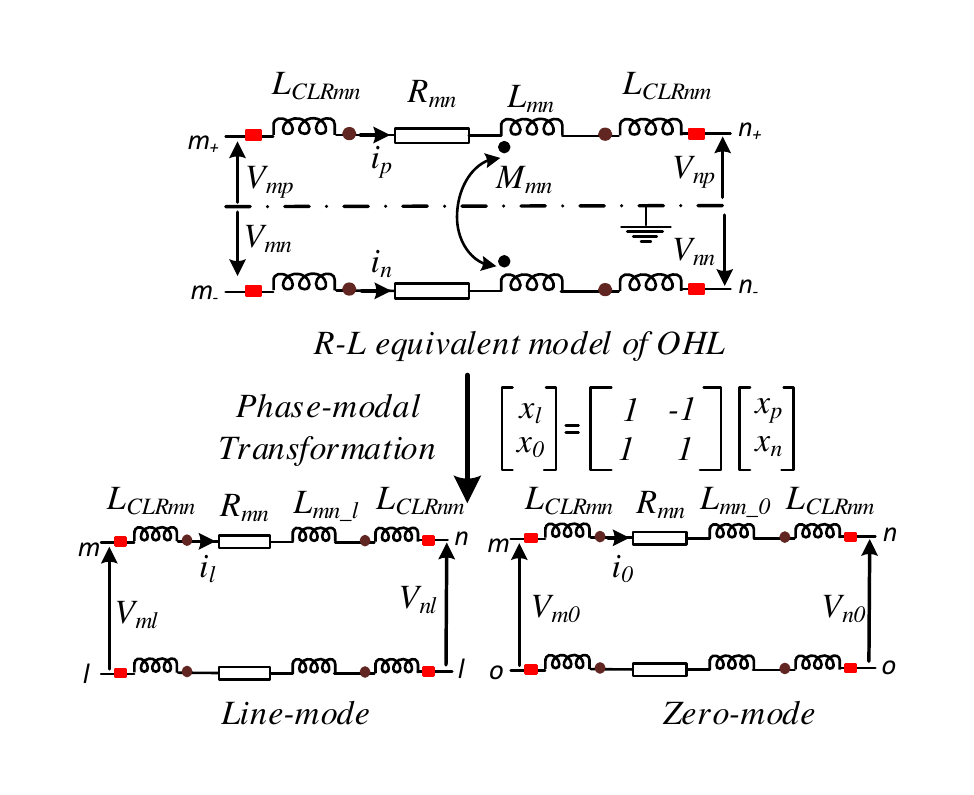}
	\caption{R-L equivalent model of OHL undergoing Phase-modal Transformation to give line-mode and zero-mode configurations}
	\label{fig:sys_model}
\end{figure}
The decoupling of dependency of poles of transmission lines under pole to ground (\textit{PTG}) faults is incorporated using phase-modal transformation [14] as shown in Fig. 3. \textit{x$_{l}$} and \textit{x$_{0}$} are defined as the line-mode and zero-mode variables whereas \textit{x$_{p}$} and \textit{x$_{n}$} are the positive-pole and negative-pole variables. Fig. 3 shows the line-mode and the zero-mode of R-L equivalent model of OHL obtained using the phase-modal transformation. \textit{L$_{mn\_l}$} and \textit{L$_{mn\_0}$} in Fig. 3 are defined as: 
\begin{subequations}
  \begin{empheq}[left=\empheqlbrace]{align}
     L_{mn\_l}=L_{mn}-M_{mn}\\
     L_{mn\_0}=L_{mn}+M_{mn}
  \end{empheq}
\end{subequations}

\begin{figure*}[!thb] 
\centering
	\includegraphics[width=0.8\paperwidth,  height= 3.3 cm]{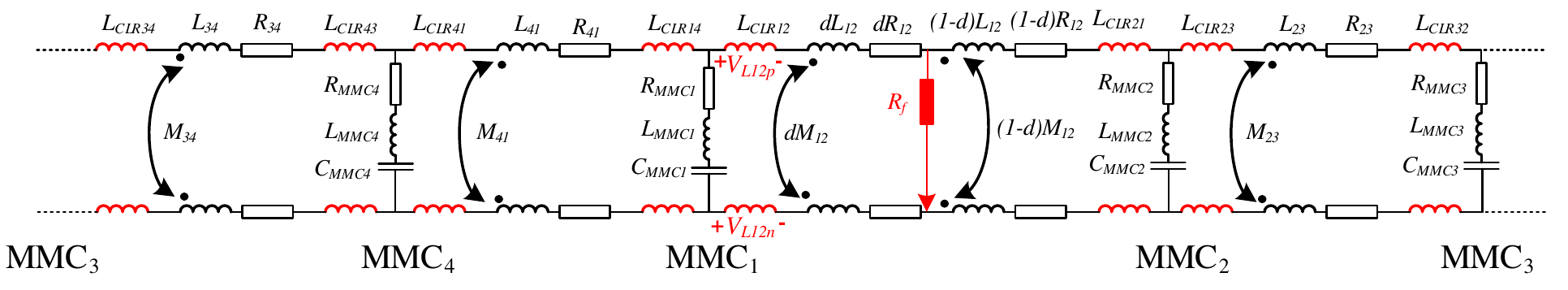}
	\caption{Equivalent network of MMC-HVDC system for a \textit{PTP} fault contingency}
	\label{fig:sys_model}
\end{figure*}
\subsection{Equivalent network during a fault contingency}
Fig. 4 shows the equivalent network of the MMC-HVDC system during a pole to pole (\textit{PTP}) fault, \textit{F$_{1}$} (see Fig. 1). The fault location, \textit{d} is the percentage of unit length for the \textit{PTP} fault with fault resistance \textit{R$_{f}$}. The simplified form of equivalent network is shown in Fig. 5 where \textit{C$_{14}$} and \textit{C$_{23}$} are the equivalent DC link terminal capacitances for line \textit{14} and \textit{23}.
\begin{figure}[!thb] 
	\includegraphics[width=1\columnwidth,  height= 3 cm]{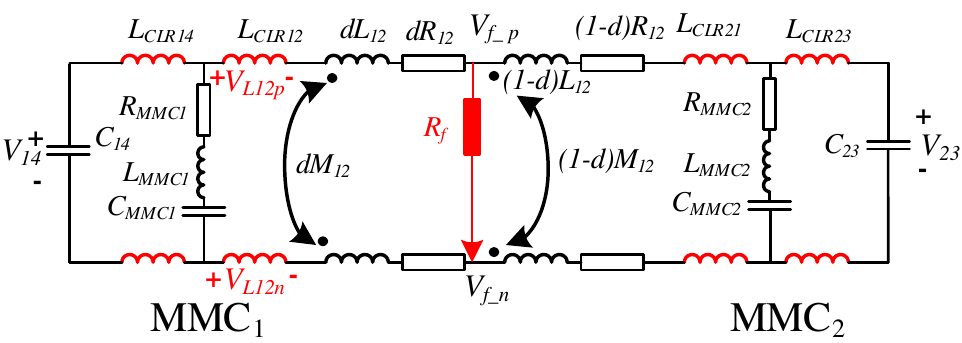}
	\caption{Simplified equivalent network of MMC-HVDC system for a \textit{PTP} fault contingency}
	\label{fig:sys_model}
\end{figure}
Using the phase-modal transformation gives the equivalent line-mode and zero-mode network for the fault, \textit{F$_{1}$}. The related networks are shown in Fig. 6.
\begin{figure}
\centering
\subcaptionbox{Line-mode equivalent network}{%
  \includegraphics[width=\columnwidth, height=3.1cm]{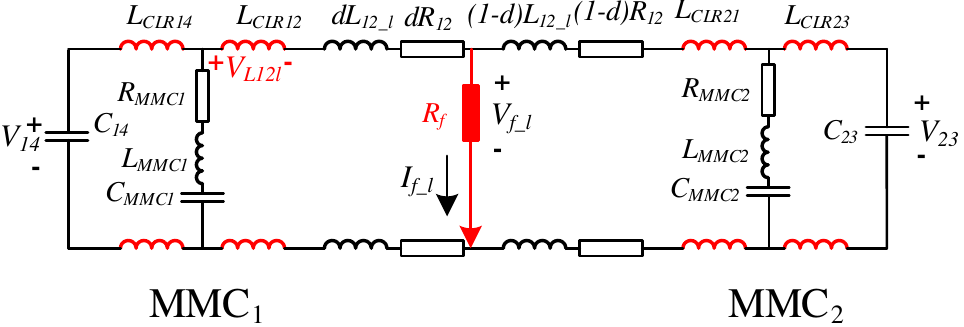}%
  }\par\medskip
\subcaptionbox{Zero-mode equivalent network}{%
  \includegraphics[width=0.9\columnwidth, height=3.1cm]{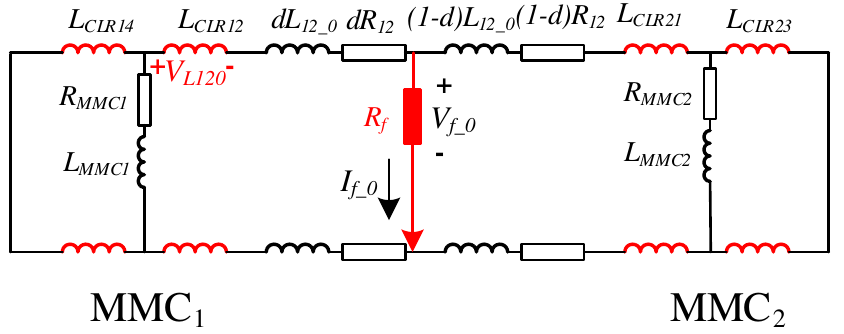}%
  }\par\medskip        
\caption{Equivalent transformed network of MMC-HVDC system for a \textit{PTP} fault contingency, \textit{F$_{1}$}}
\label{TS}
\end{figure}
\section{Fault analysis for different fault contingencies}
\subsection{Internal pole to ground (PTG) fault contingency}
For an internal positive-pole to ground (\textit{P-PTG}) fault contingency shown as \textit{F$_{1}$} in Fig. 1, \textit{V$_{f\_p}$=I$_{f\_p}$R$_{f}$} and \textit{I$_{f\_n}$=0}. Using the phase-modal transformation, the conditions are transformed to the form:
\begin{subequations}
  \begin{empheq}[left=\empheqlbrace]{align}
    & V_{f\_l}+V_{f\_0}=2I_{f\_l}R_{f}\\
    & I_{f\_l}=I_{f\_0}
  \end{empheq}
\end{subequations}
Considering equation 2, the combined equivalent mode network for an internal P-PTG is shown in Fig. 7. The idea is to derive the voltage (\textit{V$_{\scaleto{L120}{4pt}}$} and \textit{V$_{\scaleto{L121}{4pt}}$}) across the current limiting reactor (\textit{L$_{\scaleto{CLR}{4pt}}$}) for both mode networks.

\footnotesize
\begin{subequations}
  \begin{empheq}[left=\empheqlbrace]{align}
    V_{\scaleto{L120}{4pt}} &  = \begin{aligned}[t]
             &  \frac{sL_{\scaleto{CLR12}{3pt}}V_{\scaleto{dc}{4pt}}}{Z_{1}+Z_{2}||\bigg[(Z_{3}||Z_{4})+2R_{f}\bigg]}\times \frac{Z_{2}||\bigg[(Z_{3}||Z_{4})+2R_{f}\bigg]}{Z_{3}} \\
            & +\frac{sL_{\scaleto{CLR12}{3pt}}V_{\scaleto{dc}{4pt}}}{Z_{2}+Z_{1}||\bigg[(Z_{3}||Z_{4})+2R_{f}\bigg]}\times \frac{Z_{1}||\bigg[(Z_{3}||Z_{4})+2R_{f}\bigg]}{Z_{3}}
             \end{aligned}\\
     V_{\scaleto{L121}{4pt}} &  = \begin{aligned}[t]
             &  -\frac{sL_{\scaleto{CLR12}{3pt}}V_{\scaleto{dc}{4pt}}}{Z_{2}+Z_{1}||\bigg[(Z_{3}||Z_{4})+2R_{f}\bigg]}\times\frac{Z_{1}||\bigg[(Z_{3}||Z_{4})+2R_{f}\bigg]}{Z_{1}}\\
             &+\frac{sL_{\scaleto{CLR12}{3pt}}V_{\scaleto{dc}{4pt}}}{Z_{1}+Z_{2}||\bigg[(Z_{3}||Z_{4})+2R_{f}\bigg]}
             \end{aligned}
  \end{empheq}
\end{subequations}
\normalsize

where \textit{V$_{\scaleto{dc}{4pt}}$} is the DC link voltage of MMC-HVDC system. It has been assumed that each MMC is controlled at the same voltage in the system. The impedances, \textit{Z$_{1}$}-\textit{Z$_{4}$} are mathematically defined as:

\scriptsize
\begin{subequations}
  \begin{empheq}[left=\empheqlbrace]{align}
    Z_{1} &  = \begin{aligned}[t]
             &  \bigg[2sL_{\scaleto{CLR14}{3pt}}||\bigg(sL_{\scaleto{MMC1}{3pt}}+R_{\scaleto{MMC1}{3pt}}\bigg)\bigg]+2s\bigg(L_{\scaleto{CLR12}{3pt}}+dL_{\scaleto{12\_l}{4pt}}\bigg)+2dR_{\scaleto{12}{3pt}}
             \end{aligned}\\
    Z_{2} &  = \begin{aligned}[t]
             &  \bigg[2sL_{\scaleto{CLR23}{3pt}}||\bigg(sL_{\scaleto{MMC2}{3pt}}+R_{\scaleto{MMC2}{3pt}}\bigg)\bigg]+2s\bigg(L_{\scaleto{CLR21}{3pt}}+(1-d)L_{\scaleto{12\_l}{4pt}}\bigg)+2(1-d)R_{\scaleto{12}{3pt}}
             \end{aligned}\\
  Z_{3} &  = \begin{aligned}[t]
             &  \bigg[2sL_{\scaleto{CLR14}{3pt}}||\bigg(sL_{\scaleto{MMC1}{3pt}}+R_{\scaleto{MMC1}{3pt}}\bigg)\bigg]+2s\bigg(L_{\scaleto{CLR12}{3pt}}+dL_{\scaleto{12\_0}{4pt}}\bigg)+2dR_{\scaleto{12}{3pt}}
             \end{aligned}\\
    Z_{4} &  = \begin{aligned}[t]
             &  \bigg[2sL_{\scaleto{CLR23}{3pt}}||\bigg(sL_{\scaleto{MMC2}{3pt}}+R_{\scaleto{MMC2}{3pt}}\bigg)\bigg]+2s\bigg(L_{\scaleto{CLR21}{3pt}}+(1-d)L_{\scaleto{12\_0}{4pt}}\bigg)+2(1-d)R_{\scaleto{12}{3pt}}
             \end{aligned}
\end{empheq}
\end{subequations}
\normalsize

Similarly, the line-mode voltage and zero-mode voltage for Negative-pole to ground (N-PTG) are defined as:

\footnotesize
\begin{subequations}
  \begin{empheq}[left=\empheqlbrace]{align}
    V_{\scaleto{L120}{4pt}} &  = \begin{aligned}[t]
             &  -\frac{sL_{\scaleto{CLR12}{3pt}}V_{\scaleto{dc}{4pt}}}{Z_{1}+Z_{2}||\bigg[(Z_{3}||Z_{4})+2R_{f}\bigg]}\times \frac{Z_{2}||\bigg[(Z_{3}||Z_{4})+2R_{f}\bigg]}{Z_{3}} \\
            & -\frac{sL_{\scaleto{CLR12}{3pt}}V_{\scaleto{dc}{4pt}}}{Z_{2}+Z_{1}||\bigg[(Z_{3}||Z_{4})+2R_{f}\bigg]}\times \frac{Z_{1}||\bigg[(Z_{3}||Z_{4})+2R_{f}\bigg]}{Z_{3}}
             \end{aligned}\\
     V_{\scaleto{L121}{4pt}} &  = \begin{aligned}[t]
             &  -\frac{sL_{\scaleto{CLR12}{3pt}}V_{\scaleto{dc}{4pt}}}{Z_{2}+Z_{1}||\bigg[(Z_{3}||Z_{4})+2R_{f}\bigg]}\times\frac{Z_{1}||\bigg[(Z_{3}||Z_{4})+2R_{f}\bigg]}{Z_{1}}\\
             &+\frac{sL_{\scaleto{CLR12}{3pt}}V_{\scaleto{dc}{4pt}}}{Z_{1}+Z_{2}||\bigg[(Z_{3}||Z_{4})+2R_{f}\bigg]}
             \end{aligned}
  \end{empheq}
\end{subequations}
\normalsize

\begin{figure}[!thb] 
\centering
	\includegraphics[width=\columnwidth,  height= 5.4cm]{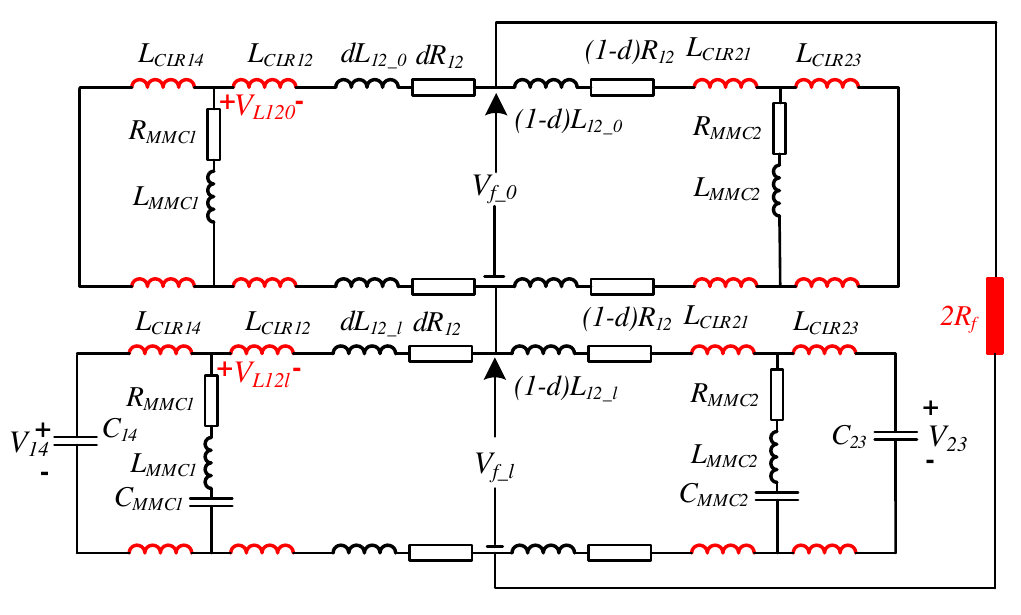}
	\caption{Combined equivalent mode network for an internal PTG fault}
	\label{fig:sys_model}
\end{figure}
\subsection{External PTG fault contingency}
\subsubsection{Backward External PTG fault contingency}
For a backward external \textit{PTG} fault contingency, \textit{F$_{3}$} (see Fig. 1), the combined equivalent mode network is shown in Fig. 8(a). The line-mode voltage and zero-mode voltage for a backward external \textit{PTG} fault are defined as:

\footnotesize
\begin{subequations}
  \begin{empheq}[left=\empheqlbrace]{align}
    V_{\scaleto{L120}{4pt}} &  = \begin{aligned}[t]
             &  -\frac{sL_{\scaleto{CLR12}{3pt}}V_{\scaleto{dc}{4pt}}}{Z_{5}+Z_{6}||\bigg[(Z_{6}||Z_{7})+2R_{f}\bigg]}\times \frac{Z_{6}||\bigg[(Z_{6}||Z_{7})+2R_{f}\bigg]}{Z_{7}} \\
            & -\frac{sL_{\scaleto{CLR12}{3pt}}V_{\scaleto{dc}{4pt}}}{Z_{6}+Z_{5}||\bigg[(Z_{6}||Z_{7})+2R_{f}\bigg]}\times \frac{Z_{5}||\bigg[(Z_{6}||Z_{7})+2R_{f}\bigg]}{Z_{7}}
             \end{aligned}\\
     V_{\scaleto{L121}{4pt}} &  = \begin{aligned}[t]
             &  \frac{sL_{\scaleto{CLR12}{3pt}}V_{\scaleto{dc}{4pt}}}{Z_{6}+Z_{5}||\bigg[(Z_{6}||Z_{7})+2R_{f}\bigg]}\times\frac{Z_{5}||\bigg[(Z_{6}||Z_{7})+2R_{f}\bigg]}{Z_{5}}\\
             &-\frac{sL_{\scaleto{CLR12}{3pt}}V_{\scaleto{dc}{4pt}}}{Z_{5}+Z_{6}||\bigg[(Z_{6}||Z_{7})+2R_{f}\bigg]}
             \end{aligned}
  \end{empheq}
\end{subequations}
\normalsize

where the impedances, \textit{Z$_{5}$}-\textit{Z$_{7}$} are defined as:

\scriptsize
\begin{subequations}
  \begin{empheq}[left=\empheqlbrace]{align}
    Z_{5} &  = \begin{aligned}[t]
             &  \bigg[2sL_{\scaleto{CLR23}{3pt}}||\bigg(sL_{\scaleto{MMC2}{3pt}}+R_{\scaleto{MMC2}{3pt}}\bigg)\bigg]+2s\bigg(L_{\scaleto{CLR12}{3pt}}+L_{\scaleto{CLR21}{3pt}}+L_{\scaleto{12\_l}{4pt}}\bigg)+2R_{\scaleto{12}{3pt}}
             \end{aligned}\\
    Z_{6} &  = \begin{aligned}[t]
             &  \bigg[2sL_{\scaleto{CLR14}{3pt}}||\bigg(sL_{\scaleto{MMC1}{3pt}}+R_{\scaleto{MMC1}{3pt}}\bigg)\bigg]
             \end{aligned}\\
  Z_{7} &  = \begin{aligned}[t]
             &  \bigg[2sL_{\scaleto{CLR23}{3pt}}||\bigg(sL_{\scaleto{MMC2}{3pt}}+R_{\scaleto{MMC2}{3pt}}\bigg)\bigg]+2s\bigg(L_{\scaleto{CLR12}{3pt}}+L_{\scaleto{CLR21}{3pt}}+L_{\scaleto{12\_0}{4pt}}\bigg)+2R_{\scaleto{12}{3pt}}
             \end{aligned}
  \end{empheq}
\end{subequations}
\normalsize

\subsubsection{Forward External PTG fault contingency}
For a forward external \textit{PTG} fault contingency, \textit{F$_{2}$} (see Fig. 1), the combined equivalent mode network is shown in Fig. 8(b). The line-mode voltage and zero-mode voltage for a forward external \textit{PTG} fault are defined as:

\footnotesize
\begin{subequations}
  \begin{empheq}[left=\empheqlbrace]{align}
    V_{\scaleto{L120}{4pt}} &  = \begin{aligned}[t]
             &  \frac{sL_{\scaleto{CLR12}{3pt}}V_{\scaleto{dc}{4pt}}}{Z_{8}+Z_{9}||\bigg[(Z_{9}||Z_{10})+2R_{f}\bigg]}\times \frac{Z_{9}||\bigg[(Z_{9}||Z_{10})+2R_{f}\bigg]}{Z_{10}} \\
            & +\frac{sL_{\scaleto{CLR12}{3pt}}V_{\scaleto{dc}{4pt}}}{Z_{9}+Z_{8}||\bigg[(Z_{9}||Z_{10})+2R_{f}\bigg]}\times \frac{Z_{8}||\bigg[(Z_{9}||Z_{10})+2R_{f}\bigg]}{Z_{10}}
             \end{aligned}\\
     V_{\scaleto{L121}{4pt}} &  = \begin{aligned}[t]
             &  -\frac{sL_{\scaleto{CLR12}{3pt}}V_{\scaleto{dc}{4pt}}}{Z_{9}+Z_{8}||\bigg[(Z_{9}||Z_{10})+2R_{f}\bigg]}\times\frac{Z_{8}||\bigg[(Z_{9}||Z_{10})+2R_{f}\bigg]}{Z_{8}}\\
             &+\frac{sL_{\scaleto{CLR12}{3pt}}V_{\scaleto{dc}{4pt}}}{Z_{8}+Z_{9}||\bigg[(Z_{9}||Z_{10})+2R_{f}\bigg]}
             \end{aligned}
  \end{empheq}
\end{subequations}
\normalsize

where the impedances, \textit{Z$_{8}$}-\textit{Z$_{10}$} are defined as:

\scriptsize
\begin{subequations}
  \begin{empheq}[left=\empheqlbrace]{align}
    Z_{8} &  = \begin{aligned}[t]
             &  \bigg[2sL_{\scaleto{CLR14}{3pt}}||\bigg(sL_{\scaleto{MMC1}{3pt}}+R_{\scaleto{MMC1}{3pt}}\bigg)\bigg]+2s\bigg(L_{\scaleto{CLR12}{3pt}}+L_{\scaleto{CLR21}{3pt}}+L_{\scaleto{12\_l}{4pt}}\bigg)+2R_{\scaleto{12}{3pt}}
             \end{aligned}\\
    Z_{9} &  = \begin{aligned}[t]
             &  \bigg[2sL_{\scaleto{CLR23}{3pt}}||\bigg(sL_{\scaleto{MMC2}{3pt}}+R_{\scaleto{MMC2}{3pt}}\bigg)\bigg]
             \end{aligned}\\
  Z_{10} &  = \begin{aligned}[t]
             &  \bigg[2sL_{\scaleto{CLR14}{3pt}}||\bigg(sL_{\scaleto{MMC1}{3pt}}+R_{\scaleto{MMC1}{3pt}}\bigg)\bigg]+2s\bigg(L_{\scaleto{CLR12}{3pt}}+L_{\scaleto{CLR21}{3pt}}+L_{\scaleto{12\_0}{4pt}}\bigg)+2R_{\scaleto{12}{3pt}}
             \end{aligned}
  \end{empheq}
\end{subequations}
\normalsize

 \begin{table}
  \centering
  \caption{Polarity of line-mode and zero-mode voltages for different \textit{PTG} faults}
  \begin{tabular}{>{\bfseries}c*{2}{c}}\toprule
    \multirow{2}{*}{\bfseries Fault Contingency} & \multicolumn{2}{c}{\bfseries Polarity} 
                                               \\\cmidrule(lr){2-3}
                       & \textit{V$_{\scaleto{L120}{4pt}}$} & \textit{V$_{\scaleto{L121}{4pt}}$}  \\ \midrule
    \textit{P-PTG} & Positive        & Positive                          \\
    \textit{N-PTG} & Negative     & Positive                          \\ 
    \textit{Backward External}     & Negative     & Negative        \\ 
    \textit{Forward External}        & Positive     & Positive\\ 
 \bottomrule
  \end{tabular}
  \end{table}

\begin{figure}
\centering
\subcaptionbox{Combined equivalent mode network for a backward external PTG fault}{%
  \includegraphics[width=\columnwidth, height=5.8cm]{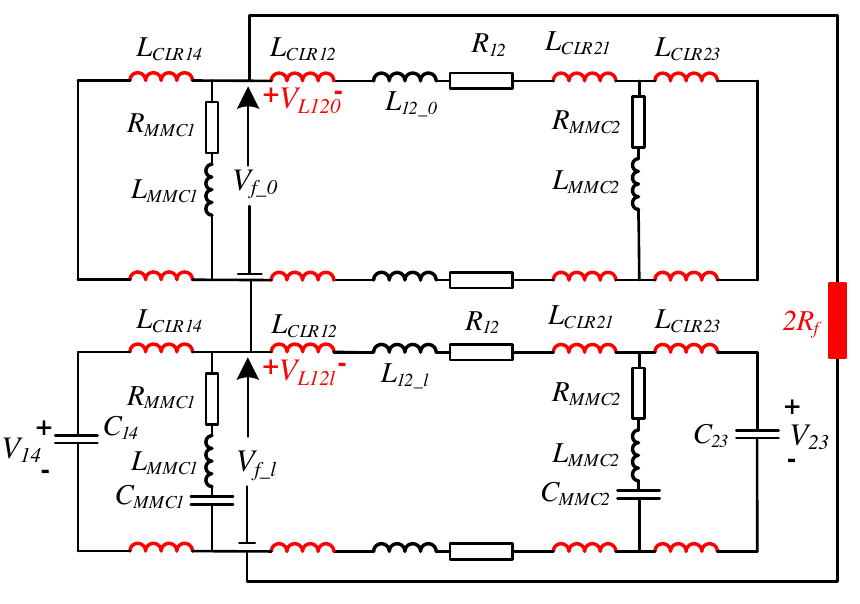}%
  }\par\medskip
\subcaptionbox{Combined equivalent mode network for a forward external PTG fault}{%
  \includegraphics[width=\columnwidth, height=5.8cm]{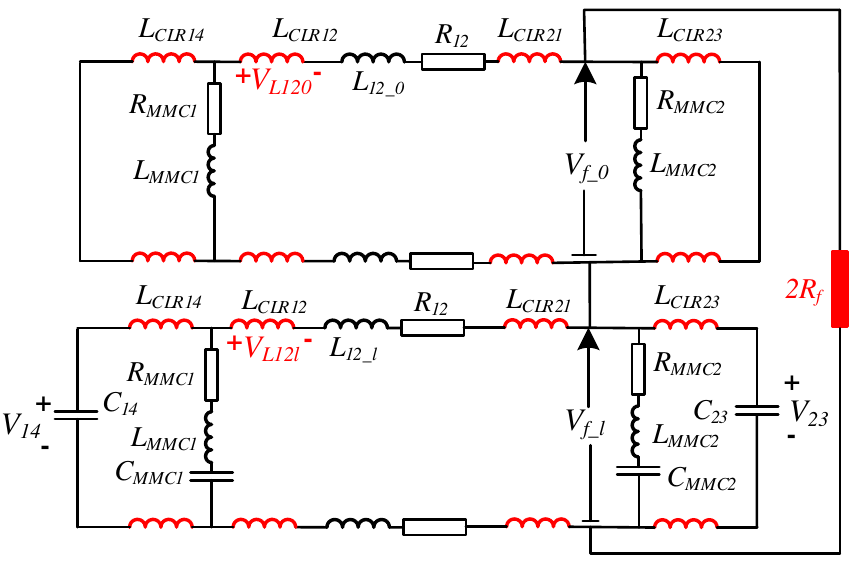}%
  }\par\medskip        
\caption{Combined equivalent mode network for external PTG fault}
\label{TS}
\end{figure}

\subsection{Internal PTP fault contingency}
For a \textit{PTP} fault contingency (F$_{1}$) in the system shown in Fig. 1, \textit{V$_{f\_p}$-V$_{f\_n}$=I$_{f\_p}$R$_{f}$} and \textit{I$_{f\_p}$+I$_{f\_n}$=0}. Applying phase-modal transformation, the conditions are transformed to the form:
\begin{subequations}
  \begin{empheq}[left=\empheqlbrace]{align}
    & V_{f\_l}=I_{f\_l}R_{f}\\
    & I_{f\_0}=0
  \end{empheq}
\end{subequations}
Considering equation 10, the equivalent mode network for an internal \textit{PTP} fault is shown in Fig. 9. The line-mode voltage and the zero-mode voltage for internal \textit{PTP} are defined as:

\footnotesize
\begin{subequations}
  \begin{empheq}[left=\empheqlbrace]{align}
    V_{\scaleto{L120}{4pt}} &  = \begin{aligned}[t]
            0
             \end{aligned}\\
     V_{\scaleto{L121}{4pt}} &  = \begin{aligned}[t]
             &  \frac{sL_{\scaleto{CLR12}{3pt}}V_{\scaleto{dc}{4pt}}}{Z_{1}+Z_{2}||R_{f}}-\frac{sL_{\scaleto{CLR12}{3pt}}V_{\scaleto{dc}{4pt}}}{Z_{2}+Z_{1}||R_{f}} \times \frac{Z_{1}||R_{f}}{Z_{1}}
             \end{aligned}
  \end{empheq}
\end{subequations}
\normalsize

\begin{figure}
\centering
\subcaptionbox{Line-mode equivalent network for an internal PTP fault}{%
  \includegraphics[width=\columnwidth, height=3cm]{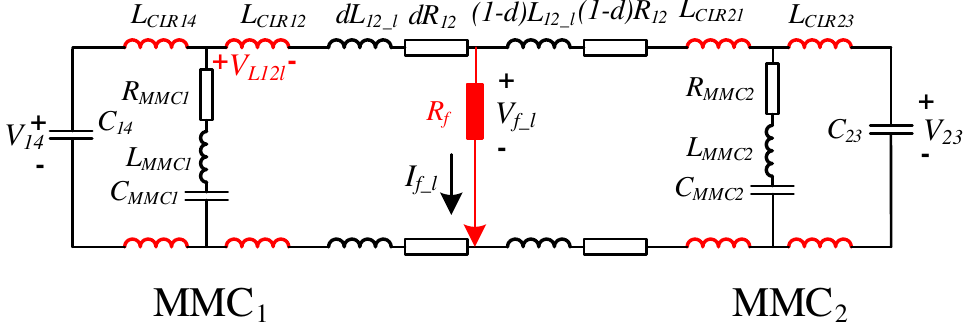}%
  }\par\medskip
\subcaptionbox{Zero-mode equivalent network for an internal PTP fault}{%
  \includegraphics[width=0.9\columnwidth, height=3cm]{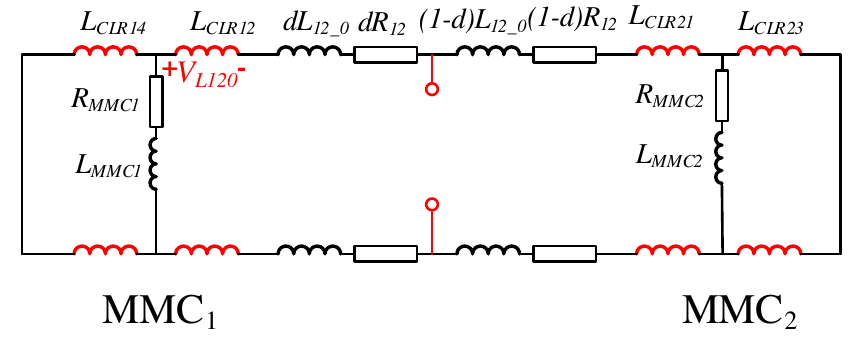}%
  }\par\medskip        
\caption{Mode Equivalent network for an internal PTP fault}
\label{TS}
\end{figure}
% \newline

\subsection{External PTP fault contingency}
Similar to the previous subsection, the line-mode voltage and the zero-mode voltage for a backward external \textit{PTP} fault are defined as:

\footnotesize
\begin{subequations}
  \begin{empheq}[left=\empheqlbrace]{align}
    V_{\scaleto{L120}{4pt}} &  = \begin{aligned}[t]
            0
             \end{aligned}\\
     V_{\scaleto{L121}{4pt}} &  = \begin{aligned}[t]
             &  -\frac{sL_{\scaleto{CLR12}{3pt}}V_{\scaleto{dc}{4pt}}}{Z_{5}+Z_{6}||R_{f}}+\frac{sL_{\scaleto{CLR12}{3pt}}V_{\scaleto{dc}{4pt}}}{Z_{6}+Z_{5}||R_{f}} \times \frac{Z_{5}||R_{f}}{Z_{5}}
             \end{aligned}
  \end{empheq}
\end{subequations}
\normalsize

Also, the line-mode voltage and the zero-mode voltage for a forward external \textit{PTP} fault are defined as:
\footnotesize
\begin{subequations}
  \begin{empheq}[left=\empheqlbrace]{align}
    V_{\scaleto{L120}{4pt}} &  = \begin{aligned}[t]
            0
             \end{aligned}\\
     V_{\scaleto{L121}{4pt}} &  = \begin{aligned}[t]
             &  \frac{sL_{\scaleto{CLR12}{3pt}}V_{\scaleto{dc}{4pt}}}{Z_{8}+Z_{9}||R_{f}}-\frac{sL_{\scaleto{CLR12}{3pt}}V_{\scaleto{dc}{4pt}}}{Z_{9}+Z_{8}||R_{f}} \times \frac{Z_{8}||R_{f}}{Z_{8}}
             \end{aligned}
  \end{empheq}
\end{subequations}
\normalsize

 \begin{table}
  \centering
  \caption{Polarity of line-mode and zero-mode voltages for different \textit{PTP} faults}
  \begin{tabular}{>{\bfseries}c*{2}{c}}\toprule
    \multirow{2}{*}{\bfseries Fault Contingency} & \multicolumn{2}{c}{\bfseries Polarity}
                                               \\\cmidrule(lr){2-3}
                       & \textit{V$_{\scaleto{L120}{4pt}}$} & \textit{V$_{\scaleto{L121}{4pt}}$}  \\ \midrule
    \textit{PTP} & 0        & Positive                          \\
    \textit{Backward External}     & 0     & Negative        \\ 
    \textit{Forward External}        & 0     & Positive\\ 
 \bottomrule
  \end{tabular}
  \end{table}

\section{Proposed Fault Identification Algorithm}
The fault identification algorithm integrates rate of change of voltage, the line-mode and zero-mode voltage analysis and the rate of change of local current to propose a reliable protection scheme for MTDC MMC-HVDC system differentiating internal and external faults and the type of fault contingency (\textit{PTP} or \textit{PTG}). In order to activate the proposed scheme, the initial condition consists of monitoring the absolute rate of change of DC link voltage of MMC terminal where the related threshold for the condition is defined as \textit{U$_{set}$}. If the condition is violated, the line-mode and zero-mode voltage parameters, \textit{V$_{\scaleto{L120}{4pt}}$} and \textit{V$_{\scaleto{L121}{4pt}}$} are evaluated. The first classification is based on the analysis of zero-mode voltage, \textit{V$_{\scaleto{L120}{4pt}}$} where \textit{PTP} and \textit{PTG} family of faults are differentiated as shown in Fig. 10. The property of zero-mode voltage, \textit{V$_{\scaleto{L120}{4pt}}$} being equal to \textit{0} for \textit{PTP} family of faults (refer to Table II) is used for the purpose. However, for the practical implementation of scheme, the threshold, \textit{e$_{set}$} is defined instead of using \textit{0} to account for external factors like noise in measurement. Once the \textit{PTP} and \textit{PTG} family of faults is classified, Table I and Table II are referred to check the conditions for further classification of fault. As evident from the tables, the polarity of line-mode and zero-mode voltages differentiate the backward external faults from the other faults (forward external and internal). 
\begin{figure}[!thb] 
\centering
	\includegraphics[width=\columnwidth,  height= 10cm]{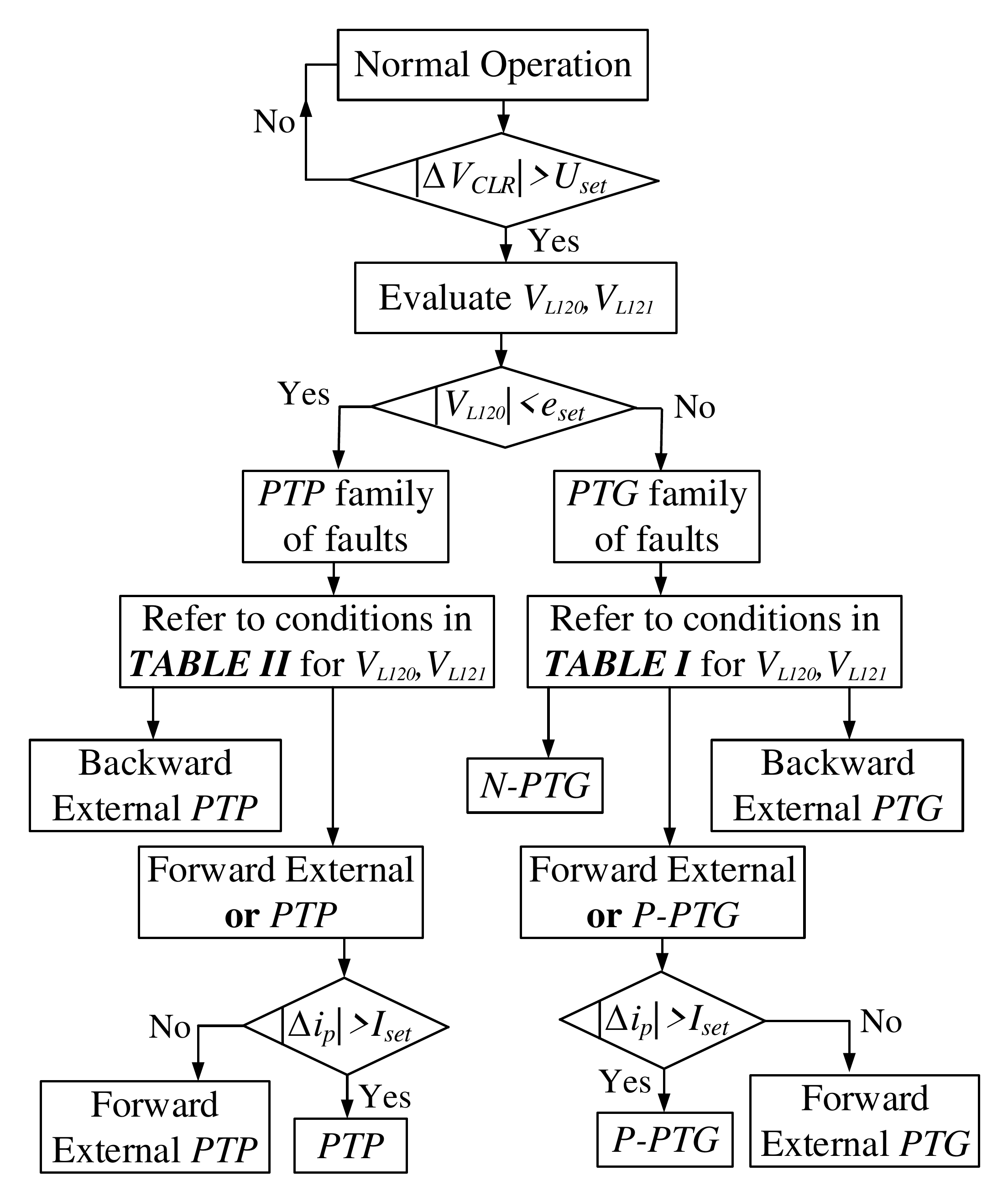}
	\caption{Flowchart of the proposed protection scheme}
	\label{fig:sys_model}
\end{figure}
  \begin{table*}
  \centering
  \caption{$|\Delta V_{CLR}|$ (kV) for different fault contingencies i.e., \textit{PTP} and \textit{PTG} at different fault location for whole range of varying current limiting reactor (CLR) values with varying fault resistances, $R_{f}$}
\begin{tabular}{@{}rrrrcrrrcrrr@{}}\toprule
& \multicolumn{3}{c}{$R_{f}=0~\Omega$} & \phantom{abc}& \multicolumn{3}{c}{$R_{f}=100~\Omega$} &
\phantom{abc} & \multicolumn{3}{c}{$R_{f}=200~\Omega$}\\
\cmidrule{2-4} \cmidrule{5-7} \cmidrule{8-12}
& $90mH$ & $130mH$& $170mH$ && $90mH$ & $130mH$& $170mH$ && $90mH$ & $130mH$& $170mH$\\ 
\midrule
\textit{PTG fault contingency}\\
10\%  & 278 & 312& 342  && 247 & 273&298 && 212 & 246&271\\
50\%  & 212& 274& 318 && 174& 218&264 && 148& 174&223\\
90\%  & 187& 227& 268 && 156& 187&204 && 124& 162&196\\
\textit{PTP fault contingency}\\
10\%  & 592 & 652& 728  && 541 & 598&656 && 492 & 552& 604\\
50\% & 481& 576& 652 && 452& 492&574 && 404& 502&562\\
90\%  & 412& 458& 504 && 376& 407&454 && 347& 467&496\\
\bottomrule
\end{tabular}
\end{table*}

However, the classification of forward external and \textit{PTP} internal or forward external and \textit{PTG} internal is not evident from the corresponding tables where the polarity is same for forward external and internal faults. Also, the unknown value of fault resistance, $R_{f}$ makes line-mode and zero-mode voltages uncertain for further classification. Hence, the absolute rate of change of local current is employed for this classification. The idea is, for a forward external fault, the rate of rise of current is limited due to larger value of current limiting reactor (L$_{\scaleto{CLR}{4pt}}$) in the path of fault current, \textit{i$_{p}$}. The threshold for the condition is defined as \textit{I$_{set}$} where the internal faults violate the threshold.

\begin{table}[h!]
\begin{center}
\centering
\caption{Parameters implemented for the validation of proposed DC fault identification scheme.}
\label{table:1}
\begin{tabular}{ c r @ {} l}
\toprule
\textbf{Parameters} & \multicolumn {2}{c}{\textbf{Value}} \\
\hline
\hline
DC link voltage \textit{(kV)} & $\pm$500\\
\hline
 Number of Sub-modules (SM) & 200   \\ 
  \hline
Capacitance of SM \textit{($\mu$F)} & 1.5$\times$10$^{3}$ \\
  \hline
Current Limiting Reactor (\textit{mH}) & 90-170 \\
  \hline
Arm Inductance (\textit{mH}) & 100 \\
  \hline
  Arm resistance (\textit{$\Omega$}) & 0.85 \\
  \hline
  Line resistance per unit (\textit{$\Omega$/m}) & 4.116$\times$10$^{-5}$ \\
  \hline
Line inductance per unit (\textit{mH/m}) & 1.256$\times$10$^{-5}$&\\
\bottomrule
\end{tabular}
\end{center}
\vspace{-8mm}
\end{table}
\section{Results and Validation}
Fig. 1 shows the four-terminal MMC-HVDC system with DC voltage of $\pm$500kV. MMCs at bus 2 and bus 3 operate in DC voltage control and constant reactive power control [23]. MMCs at bus 1 and  bus 4 operate in constant active and reactive power control. The parameters employed for the simulation validation are mentioned in Table IV. The wavelet transform is set as 1-level
and the corresponding frequency is 5 kHz–10 kHz [13]. The mother
wavelet is selected as sym8 [15]. The test system is employed using PSCAD/EMTDC based electromagnetic transient simulations.

\subsection{Evaluating trigger event threshold, \textit{U$_{set}$}}
The absolute rate of change of CLR voltage, \textit{V$_{\scaleto{CLR}{4pt}}$} is used as the trigger event for the evaluation of decisive parameters, \textit{V$_{\scaleto{L120}{4pt}}$} and \textit{V$_{\scaleto{L121}{4pt}}$}. The threshold for the trigger event, \textit{U$_{set}$} is defined for a DC fault contingency in the system with the objective that normal system transients should not be identified as a fault. Also, considering that the trigger should be activated in time keeping margin for the complete algorithm. The decision for \textit{U$_{set}$} is taken considering different type of fault contingencies, \textit{PTP} and \textit{PTG} with varying fault location in the OHL for varying fault resistance, \textit{R$_{f}$} (see Table III). The study also includes a range of values for CLR. As shown in Table III, the magnitude of \textit{|$\Delta$V$_{\scaleto{CLR}{4pt}}$|} decreases with the increase in fault resistance, R$_{f}$ and increase in the fault location percentage in a line. However, with increasing value of CLR, \textit{|$\Delta$V$_{\scaleto{CLR}{4pt}}$|} increases. The minimum value of \textit{|$\Delta$V$_{\scaleto{CLR}{4pt}}$|} is realised for a \textit{PTG} fault contingency at a fault location, \textit{100$\%$} of the length of OHL for a CLR value of \textit{90mH} and a fault resistance of \textit{200$\Omega$}. Based on the results shown in Table III, \textit{U$_{set}$} is taken to be \textit{100kV} keeping a considerable safety margin.
\begin{figure}[!thb] 
\centering
	\includegraphics[width=\columnwidth,  height= 9cm]{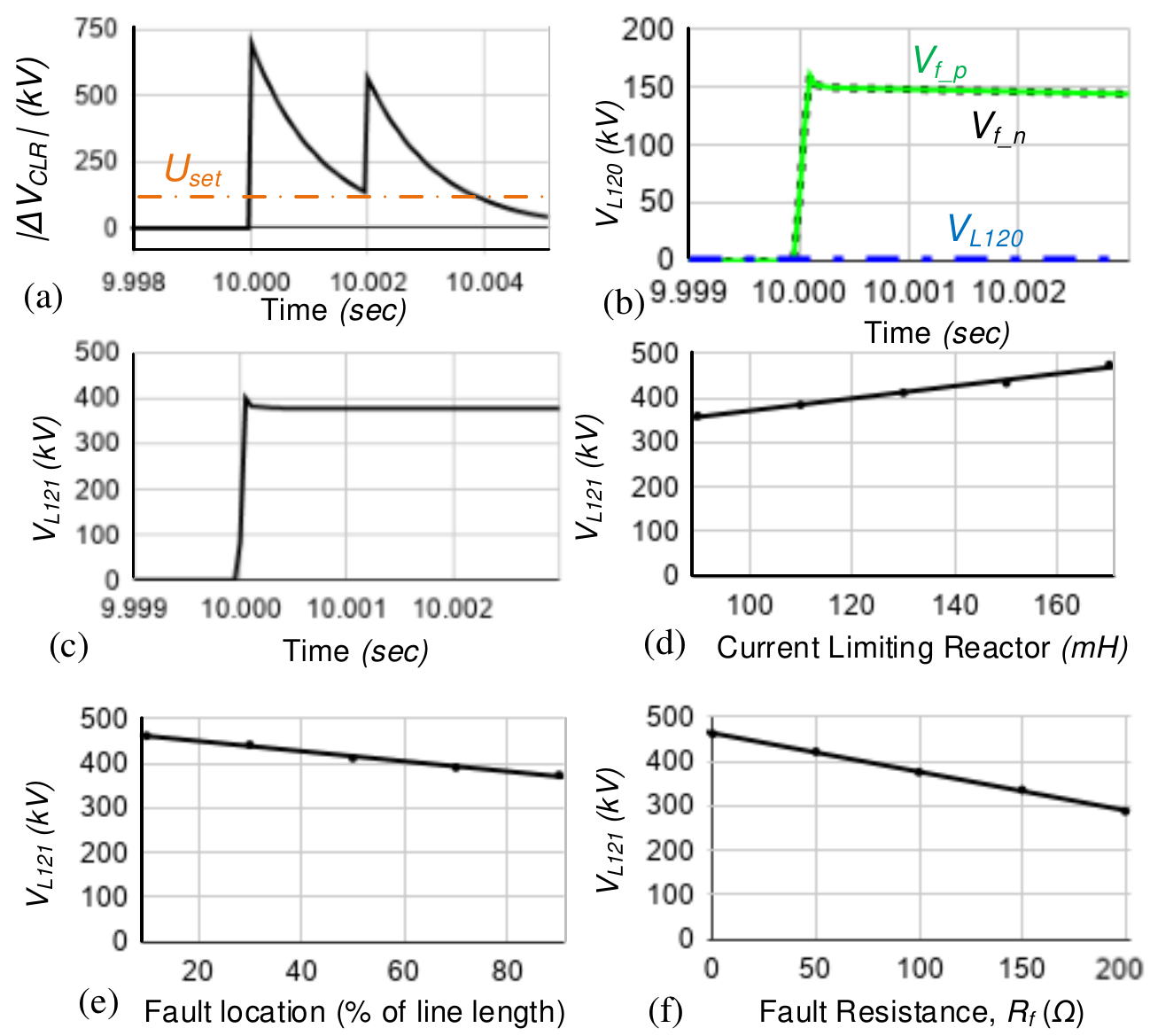}
	\caption{(a)\textit{|$\Delta$V$_{\scaleto{CLR}{4pt}}$|} (kV) vs time (sec); (b) \textit{V$_{\scaleto{f\_p}{4pt}}$} (green), \textit{V$_{\scaleto{f\_n}{4pt}}$} (black) and \textit{V$_{\scaleto{L120}{4pt}}$} (blue) vs time (sec); (c)  
	\textit{V$_{\scaleto{L121}{4pt}}$} (kV) vs time (sec); (d) \textit{V$_{\scaleto{L121}{4pt}}$} (kV) vs CLR (mH); (e) \textit{V$_{\scaleto{L121}{4pt}}$} (kV) vs Fault location ($\%$); (f) \textit{V$_{\scaleto{L121}{4pt}}$} (kV) vs Fault Resistance ($\Omega$)}
	\label{fig:sys_model}
\end{figure}
\subsection{Validation for \textit{PTP} fault contingency}
The protection scheme is tested for a \textit{PTP} fault contingency in the system at \textit{t=10 s}. Fig. 11(a) shows \textit{|$\Delta$V$_{\scaleto{CLR}{4pt}}$|} plot with the variation of time. The threshold, \textit{U$_{set}$} is violated almost instantaneously. Fig. 11(b) shows \textit{V$_{f\_p}$} (green) and \textit{V$_{f\_n}$} (dotted black) to finally give V$_{\scaleto{L120}{4pt}}$ (dotted blue) which coincides with the \textit{0} line validating the idea of zero-mode voltage for a \textit{PTP} equal to \textit{0}. Fig. 11(c) shows V$_{\scaleto{L121}{4pt}}$ variation with time whereas Fig. 11(d) shows V$_{\scaleto{L121}{4pt}}$ with the variation of CLR. It can be seen that the line-mode voltage increases with the increase in value of CLR. Fig. 11(e) and 11(f) shows the variation of line-mode voltage with fault location and fault resistance in a system. The line-mode voltage decreases with the increase in fault location or fault resistance in a system.
\subsection{Evaluating fault type distinguishing threshold, \textit{e$_{set}$}}
The lowest value of zero-mode voltage for \textit{PTG} fault (\textit{V$_{\scaleto{L120}{4pt}}$=30kV}) is recorded for a fault at \textit{100$\%$} of line length with fault resistance of \textit{200$\Omega$} and \textit{90mH} as the CLR (maximum fault location, maximum fault resistance and minimum CLR considered in the analysis as explained in section V(A)). Hence, the threshold \textit{e$_{set}$}, which is used to distinguish between the type of fault contingency (\textit{PTP} or \textit{PTG}) (see Fig. 10) is taken considering nature of zero-mode voltage in both cases. For practical implementation, a safety factor of 1/3 is used to implement \textit{e$_{set}$} and is set as \textit{10 kV}. This means that any fault with zero-mode voltage greater than \textit{10 kV} is considered as a \textit{PTG} fault whereas the fault is \textit{PTP} otherwise. 
\subsection{Validation for \textit{PTG} fault contingency}
The scheme is also validated for a \textit{PTG} fault with fault resistance, \textit{R$_{f}$=100$\Omega$} in the system at \textit{t=10 s}. Fig. 12(a) shows \textit{|$\Delta$V$_{\scaleto{CLR}{4pt}}$|} plot with the variation of time whereas Fig. 12(b) shows V$_{\scaleto{L120}{4pt}}$ for a \textit{PTG} fault. Fig. 12(c) shows the line-mode voltage for a \textit{PTG} fault. 
\begin{figure}[!thb] 
\centering
	\includegraphics[width=\columnwidth,  height= 6cm]{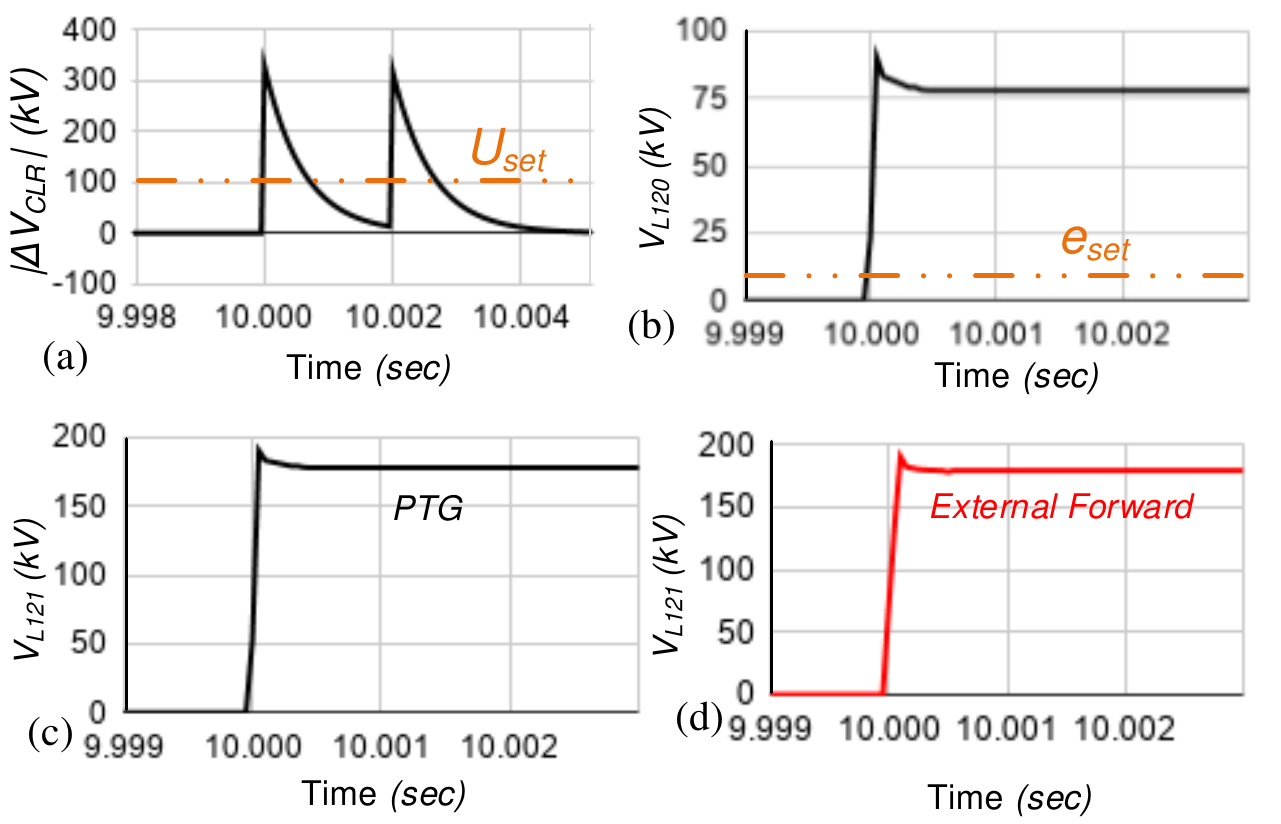}
\caption{(a)\textit{|$\Delta$V$_{\scaleto{CLR}{4pt}}$|} (kV) vs time (sec); (b) \textit{V$_{\scaleto{L120}{4pt}}$} vs time (sec); (c)\textit{V$_{\scaleto{L121}{4pt}}$} (kV) vs time (sec) for \textit{PTP}; (d) \textit{V$_{\scaleto{L121}{4pt}}$} (kV) vs time (sec) for external forward fault}
	\label{fig:sys_model}
\end{figure}
Fig. 12(d) shows line-mode voltage for a bolted external forward fault in the system. It is observed that Fig. 12(d) is identical to Fig. 12(c) considering the magnitude of line-mode voltage. Since, the system subjected to a fault has an unknown value of fault location and fault resistance, it is difficult to distinguish an internal fault and an external forward fault considering only the magnitude of line-mode voltage as proposed in [14]. The absolute rate of local current is employed to distinguish between an internal fault and an external forward fault in a system as explained subsequently.
\newline
The idea is the absolute rate of change of local current is dependent on the inductance in the path of fault current. For the case of external forward fault contingency, the minimum equivalent inductance in the path is \textit{L$_{\scaleto{CLR12}{3pt}}$+L$_{\scaleto{CLR21}{3pt}}$+L$_{12\_l}$} whereas the maximum possible inductance in the path (at \textit{100$\%$} fault location) for an internal fault contingency is \textit{L$_{\scaleto{CLR12}{3pt}}$+L$_{12\_l}$}.
\begin{figure}[!thb] 
\centering
	\includegraphics[width=\columnwidth,  height= 6cm]{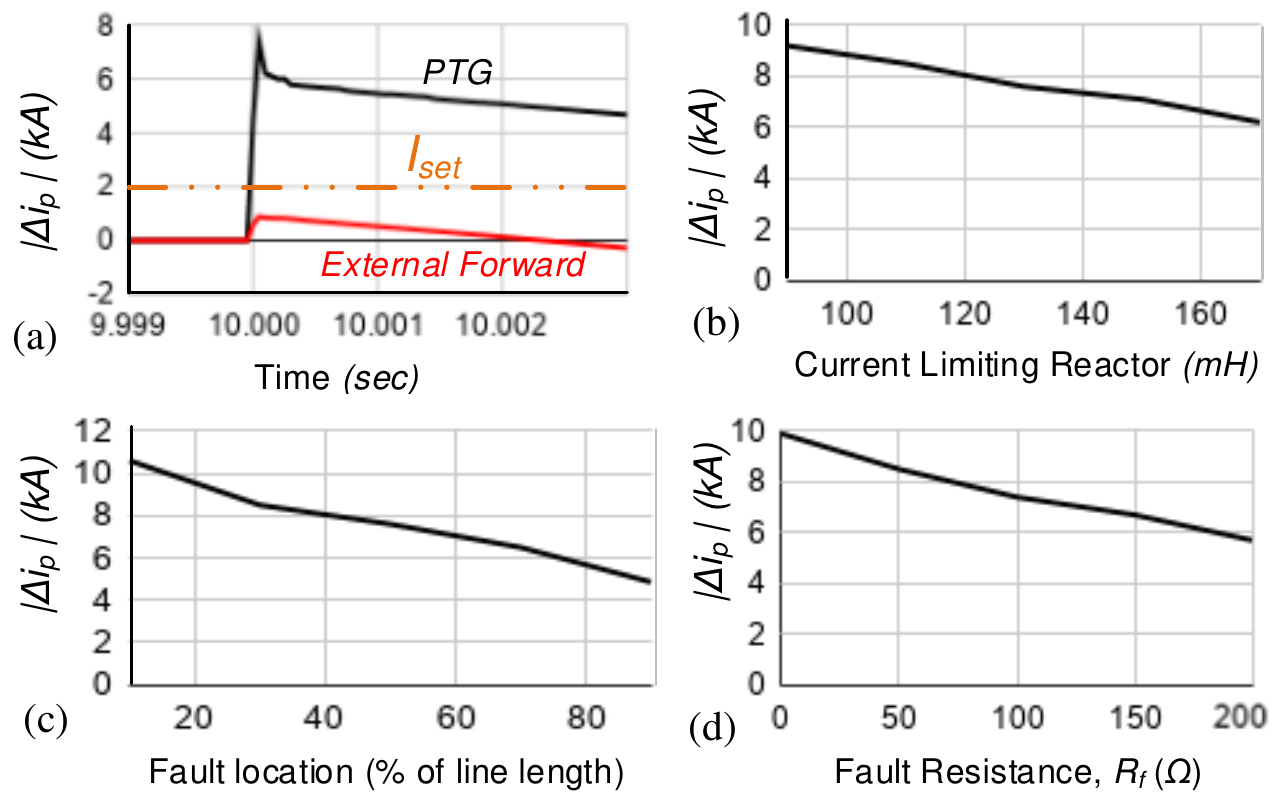}
	\caption{(a)\textit{|$\Delta$i$_{\scaleto{p}{4pt}}$|} (kA) vs time (sec) for \textit{PTG} (black), external forward (red); (b) \textit{|$\Delta$i$_{\scaleto{p}{4pt}}$|} (kA) vs CLR (mH); (c)\textit{|$\Delta$i$_{\scaleto{p}{4pt}}$|} (kA) vs Fault location ($\%$); (d) \textit{|$\Delta$i$_{\scaleto{p}{4pt}}$|} (kA) vs Fault Resistance ($\Omega$)}
	\label{fig:sys_model}
\end{figure}
Fig. 13(a) shows absolute rate of change of local current for a \textit{PTG} fault and also for an external forward fault. The threshold is decided considering the subsequent variation of  \textit{|$\Delta$i$_{\scaleto{p}{4pt}}$|} with the variation of CLR, fault location and fault resistance as shown in Fig. 13(b)-13(d). The parameter, \textit{|$\Delta$i$_{\scaleto{p}{4pt}}$|} decreases with the increase in CLR, fault location and fault resistance as evident from Fig. 13. The threshold, \textit{I$_{set}$} is taken to be \textit{2 kA} considering the above variation.

\subsection{ Effect of White Gaussian Noise (WGN)}
White Gaussian Noise (WGN) in the measurement induce high frequency components in the measured data of current and voltage. This can cause inaccuracy in the results for the fault identification of the proposed scheme. As a result, a rolling mean filter is employed with a moving window of 50 sample steps [16]. WGN is random in nature with the property of zero mean [14]. Hence, the filter eliminates the effect of WGN to a great extent without affecting the accuracy of the proposed fault identification scheme. Fig. 14 shows \textit{|$\Delta$i$_{\scaleto{p}{4pt}}$|} and \textit{|$\Delta$V$_{\scaleto{CLR}{4pt}}$|} in the presence of WGN of 30dB. It can be seen that the scheme works fine violating \textit{I$_{set}$} and \textit{U$_{set}$} respectively. 
\begin{figure}[!thb] 
\centering
	\includegraphics[width=\columnwidth,  height= 3cm]{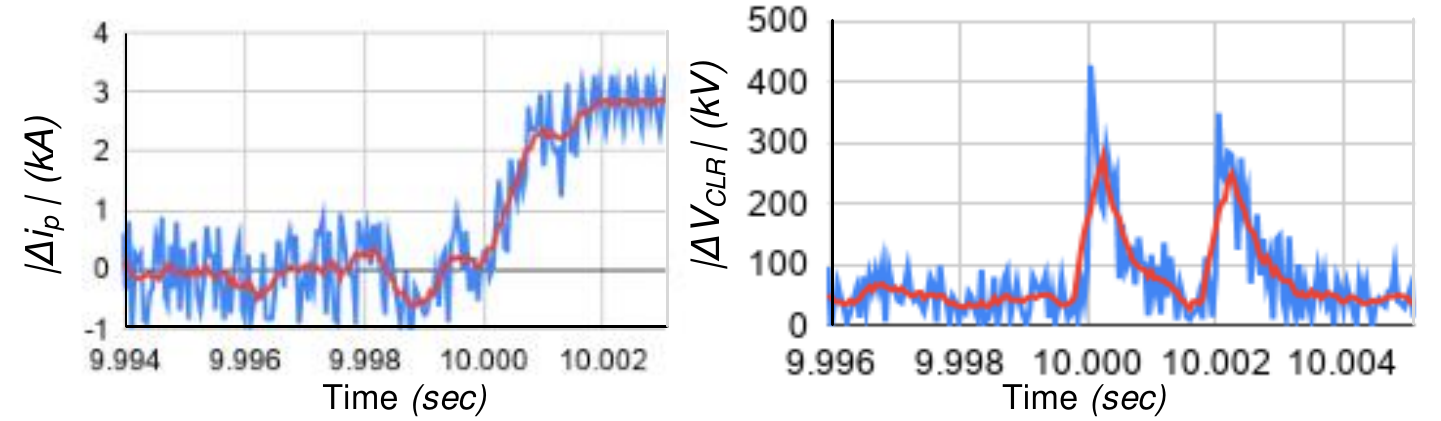}
	\caption{(a) \textit{|$\Delta$i$_{\scaleto{p}{4pt}}$|} (kA) vs time (sec) in the presence of WGN of 30dB; (b) \textit{|$\Delta$V$_{\scaleto{CLR}{4pt}}$|} (kV) vs time (sec) in the presence of WGN of 30dB }
	\label{fig:sys_model}
\end{figure}
\section{Conclusion}
 The contribution of the proposed work has been to address the issue of selectivity and dependability for localised protection based DC fault identification in multi-terminal MMC-HVDC systems. The proposed scheme takes equivalent network of system employing phase-modal transformation and analysing the line-mode and zero-mode voltage across current limiting reactor (CLR) for different possible contingencies in the presence of fault resistance. The proposed protection scheme integrates absolute rate of change of CLR voltage, line-mode and zero mode voltage and absolute rate of change of local current. This ensures the scheme distinguishes between an internal fault and an external forward fault in a system. The scheme works well for high impedance faults (HIFs) and in the presence of white gaussian noise (WGN).


\begin{thebibliography}{20}


\bibitem{}
North Sea Wind Power Hub - Consortium Partners. (2019) Power hub as an island. [Online]. https://northseawindpowerhub.eu/wp-content/uploads/2019/07/NSWPH-Benefit-study-for-potential-locations-of-an-offshore-hub-island-1.pdf
\bibitem{}
A. Orths, A. Hiorns, R. van Houtert, L. Fisher, and C. Fourment, “The european north seas countries’ offshore grid initiative — the way forward,” in 2012 IEEE Power and Energy Society General Meeting, July 2012, pp. 1–8.
\bibitem{}
W. Xiang, S. Yang, L. Xu, J. Zhang, W. Lin and J. Wen, "A Transient Voltage-Based DC Fault Line Protection Scheme for MMC-Based DC Grid Embedding DC Breakers," in IEEE Transactions on Power Delivery, vol. 34, no. 1, pp. 334-345, Feb. 2019.
\bibitem{}
G. Liu, F. Xu, Z. Xu, Z. Zhang and G. Tang, "Assembly HVDC Breaker for HVDC Grids With Modular Multilevel Converters," in IEEE Transactions on Power Electronics, vol. 32, no. 2, pp. 931-941, Feb. 2017, doi: 10.1109/TPEL.2016.2540808.
\bibitem{}
X. Han, W. Sima, M. Yang, L. Li, T. Yuan and Y. Si, "Transient Characteristics Under Ground and Short-Circuit Faults in a ${\pm \text{500}\,\text{kV}}$ MMC-Based HVDC System With Hybrid DC Circuit Breakers," in IEEE Transactions on Power Delivery, vol. 33, no. 3, pp. 1378-1387, June 2018, doi: 10.1109/TPWRD.2018.2795800.
\bibitem{} 
M. N. Haleem and A. D. Rajapakse, "Fault Type Discrimination in HVDC Transmission Lines Using Rate of Change of Local Currents," in IEEE Transactions on Power Delivery.
doi: 10.1109/TPWRD.2019.2922944
\bibitem{} 
J. Sneath and A. D. Rajapakse, "Fault Detection and Interruption in an Earthed HVDC Grid Using ROCOV and Hybrid DC Breakers," in IEEE Transactions on Power Delivery, vol. 31, no. 3, pp. 973-981, June 2016.
\bibitem{} 
R. Li, L. Xu and L. Yao, "DC Fault Detection and Location in Meshed Multiterminal HVDC Systems Based on DC Reactor Voltage Change Rate," in IEEE Transactions on Power Delivery, vol. 32, no. 3, pp. 1516-1526, June 2017.
\bibitem{} 
J. Liu, N. Tai and C. Fan, "Transient-Voltage-Based Protection Scheme for DC Line Faults in the Multiterminal VSC-HVDC System," in IEEE Transactions on Power Delivery, vol. 32, no. 3, pp. 1483-1494, June 2017.
\bibitem{} 
Z. Dai, N. Liu, C. Zhang, X. Pan and J. Wang, "A Pilot Protection for HVDC Transmission Lines Based on Transient Energy Ratio of DC Filter Link," in IEEE Transactions on Power Delivery.
doi: 10.1109/TPWRD.2019.2950350
\bibitem{} 
S. Li, W. Chen, X. Yin, D. Chen and Y. Teng, "A Novel Integrated Protection for VSC-HVDC Transmission Line Based on Current Limiting Reactor Power," in IEEE Transactions on Power Delivery.
doi: 10.1109/TPWRD.2019.2945412
\bibitem{}
C. Li, A. M. Gole and C. Zhao, "A Fast DC Fault Detection Method Using DC Reactor Voltages in HVdc Grids," in IEEE Transactions on Power Delivery, vol. 33, no. 5, pp. 2254-2264, Oct. 2018.
\bibitem{} 
W. Xiang, S. Yang, L. Xu, J. Zhang, W. Lin and J. Wen, "A Transient Voltage-Based DC Fault Line Protection Scheme for MMC-Based DC Grid Embedding DC Breakers," in IEEE Transactions on Power Delivery, vol. 34, no. 1, pp. 334-345, Feb. 2019.
doi: 10.1109/TPWRD.2018.2874817
\bibitem{}
S. Yang, W. Xiang, R. Li, X. Lu, W. Zuo and J. Wen, "An Improved DC fault Protection Algorithm for MMC HVDC Grids based on Modal Domain Analysis," in IEEE Journal of Emerging and Selected Topics in Power Electronics.
doi: 10.1109/JESTPE.2019.2945200
\bibitem{}
Y. M. Yeap, N. Geddada, and A. Ukil, “Analysis and validation of wavelet transform based DC fault detection in HVDC system,” in Appl. Soft Comput., vol. 61, pp. 127–137, 2017
\bibitem{}
V. Nougain, V. Nougain and S. Mishra, "Low-voltage DC ring-bus microgrid protection with rolling mean technique," 2018 IEEMA Engineer Infinite Conference (eTechNxT), New Delhi, 2018, pp. 1-6.
\bibitem{}
J. Xu, C. Zhao, Y. Xiong, C. Li, Y. Ji and T. An, "Optimal Design of MMC Levels for Electromagnetic Transient Studies of MMC-HVDC," in IEEE Transactions on Power Delivery, vol. 31, no. 4, pp. 1663-1672, Aug. 2016.
\bibitem{}
E. Kontos, R. T. Pinto, S. Rodrigues and P. Bauer, "Impact of HVDC Transmission System Topology on Multiterminal DC Network Faults," in IEEE Transactions on Power Delivery, vol. 30, no. 2, pp. 844-852, April 2015, doi: 10.1109/TPWRD.2014.2357056.
\bibitem{}
C. Li, C. Zhao, J. Xu, Y. Ji, F. Zhang and T. An, "A Pole-to-Pole Short-Circuit Fault Current Calculation Method for DC Grids," in IEEE Transactions on Power Systems, vol. 32, no. 6, pp. 4943-4953, Nov. 2017, doi: 10.1109/TPWRS.2017.2682110.
\bibitem{} 
P. T. Lewis, B. M. Grainger, H. A. Al Hassan, A. Barchowsky and G. F. Reed, "Fault Section Identification Protection Algorithm for Modular Multilevel Converter-Based High Voltage DC With a Hybrid Transmission Corridor," in IEEE Transactions on Industrial Electronics, vol. 63, no. 9, pp. 5652-5662, Sept. 2016.
\bibitem{} 
X. Li, Q. Song, W. Liu, H. Rao, S. Xu and L. Li, "Protection of Nonpermanent Faults on DC Overhead Lines in MMC-Based HVDC Systems," in IEEE Transactions on Power Delivery, vol. 28, no. 1, pp. 483-490, Jan. 2013.
\bibitem{} 
J. Yang, J. E. Fletcher and J. O'Reilly, "Short-Circuit and Ground Fault Analyses and Location in VSC-Based DC Network Cables," in IEEE Transactions on Industrial Electronics, vol. 59, no. 10, pp. 3827-3837, Oct. 2012.
\bibitem{} 
Y. Chen, S. Zhao, Z. Li, X. Wei and Y. Kang, "Modeling and Control of the Isolated DC–DC Modular Multilevel Converter for Electric Ship Medium Voltage Direct Current Power System," in IEEE Journal of Emerging and Selected Topics in Power Electronics, vol. 5, no. 1, pp. 124-139, March 2017.


\end{thebibliography}
\end{document}